\documentclass[12pt]{iopart}

\usepackage{graphicx}
\usepackage{epstopdf}

\begin{document}

\title[IBM for A$\sim$100 nuclei]
{Description of nuclei in the A$\sim$100 mass region
with the interacting boson model}

\author{M~B\"oy\"ukata$^1$, P~Van~Isacker$^2$ and \.{I}~Uluer$^1$}

\address{
$^1$Physics Department, Faculty of Science and Arts, University of K\i r\i kkale,\\
71100 K\i r\i kkale, Turkey}
\address{
$^2$Grand Acc\'el\'erateur National d'Ions Lourds, CEA/DSM--CNRS/IN2P3, B.P.~55027,\\
F-14076 Caen Cedex 5, France}


\begin{abstract}
Even--even nuclei in the $A\sim100$ mass region
are investigated within the framework of the interacting boson model-1 (\mbox{IBM-1}).
The study includes energy spectra
and electric quadrupole transition properties
of zirconium, molybdenum, ruthenium and palladium isotopes
with neutron number $N\geq52$.
A global parametrization of the \mbox{IBM-1} Hamiltonian is found
leading to a description of about 300 collective levels in 30 nuclei
with a root-mean-square deviation from the observed level energies of 120~keV.
The importance of the $d_{5/2}$ subshell closure
at neutron number $N=56$ is pointed out.
The geometric character of the nuclei can be visualized
by plotting the potential energy surface $V(\beta,\gamma)$
obtained from the \mbox{IBM-1} Hamiltonian in the classical limit.
The parametrization established on the basis of known elements
is used to predict properties of the unknown, neutron-rich isotopes
$^{106}$Zr, $^{112}$Mo, $^{116}$Ru and $^{122}$Pd.
\end{abstract}

\vspace{2pc}
\maketitle

\section{Introduction}
\label{s_intro}
The production and use of radioactive beams
is a rapidly developing new field in nuclear physics.
Pioneering experiments are taking place,
dedicated facilities are commissioned
and new ones are planned.
The primary aim of this research activity
is the study of so-called exotic nuclei, that is,
short-lived nuclei far removed from the line of stability
because of an unusual ratio of numbers of neutrons $N$ over protons $Z$.
Parallel with the rapid developments
of experimental techniques to observe exotic nuclei,
there has been a surge in the interest
in their theoretical description.

The main motivation for these studies is the possibility
that wide variations in $N/Z$
may have a profound influence on shell structure,
which constitutes the basis of our understanding of nuclei.
There exist currently two standard theoretical approaches
for the description of exotic nuclei,
namely energy density functional theory~\cite{Bender03}
and the nuclear shell model~\cite{Caurier05}.
In this paper we explore the use of a third approach,
namely the interacting boson model (IBM)
of Arima and Iachello~\cite{Arima76,Arima78,Arima79},
to predict properties of exotic nuclei.

The IBM is a semi-microscopic model of nuclei,
positioned intermediately between single-particle and collective models.
The model contains a vibrational and a rotational limit
(as well as one which can be considered as intermediate)---which
connects well with the phenomenology of nuclei---and
can be brought into relation with the nuclear shell model.
In the original version of the IBM, applicable to even--even nuclei,
the basic building blocks are $s$ and $d$ bosons
with angular momenta $L=0$ and $L=2$, respectively.
The $s$ and $d$ bosons can be interpreted
as correlated Cooper pairs formed by two nucleons in the valence shell
coupled to angular momenta $L=0$ and $L=2$.
This interpretation constitutes the basis of the connection
between the IBM and the nuclear shell model,
and leads in a natural way to a version of the model
where a distinction is made between neutron and proton bosons,
the so-called \mbox{IBM-2}~\cite{Arima77}.

If one restricts oneself to configurations
that are symmetric in the neutron and proton bosons,
the simplest version of the model arises, known as~\mbox{IBM-1}.
The standard \mbox{IBM-1} Hamiltonian
has 1 single-particle energy and 5 two-body interactions
if only relative (excitation) energies are considered,
increasing to $2+7$ if absolute binding energies
are included in the analysis.
These coefficients enter as free parameters in the model
and must be adjusted to the data.
However, because of the connection with the nuclear shell model,
the dependence of the energies and interactions
on the number of valence neutrons and protons is known qualitatively~\cite{Otsuka78}.
The long-term aim of this work is to develop a systematic \mbox{IBM-1} parametrization
(that is, to determine the functional dependence of the various coefficients
on the number of valence neutrons and protons) for entire regions of nuclei,
and to do so on the basis of a fit to the data, guided by microscopic arguments.
Once the appropriate functional dependence has been determined,
it will be possible to make spectroscopic predictions
by extrapolation to unknown regions of nuclei.
The predictive power of this procedure
will depend crucially on the counting of the valence nucleons
and hence on the shell structure of nuclei
and its evolution in exotic regions.

The idea to use the IBM to extrapolate
from known to unknown regions of the nuclear chart is not new.
In fact, in the late 1970s and early 1980s,
long before any talk of exotic nuclei,
this technique was applied on many {\em isotope} series
in the context of the \mbox{IBM-2}.
For a review of the literature, see Iachello and Arima~\cite{Iachello87}.
Occasionally also {\em isotone} series have been considered~\cite{Gomez95}
and recently also a combination
of {\em isobars} and {\em isotones}~\cite{Lalkovski09}.
However, to our knowledge a comprehensive and systematic description
with the IBM of entire {\em regions} of nuclei has not yet been tried.
The closest related study is due to Casten {\it et al.}~\cite{Casten85b}
who elaborated a multi-nucleus \mbox{IBM-1} parametrization
based on the $N_\nu N_\pi$ scheme~\cite{Casten85}.
Our intention is to develop this idea further
with more refined Hamiltonians
that better reproduce observed nuclear properties.
We also mention that our investigation has similar aims
as those of the recent work of the Tokyo group~\cite{Nomura08},
({\it i.e.}, the use of the IBM in the description of exotic nuclei);
in the latter studies the connection of the IBM with mean-field models
is exploited to arrive at a suitable parametrization
while here the long-term aim is to draw inspiration
from its connection with the shell model.

In this paper we present a first calculation of this type
and apply the \mbox{IBM-1} to an entire region of the nuclear mass chart,
namely the even--even Zr, Mo, Ru and Pd nuclei with $A\sim100$.
Section~\ref{s_param}
gives a description of the parametrization of the Hamiltonian used.
The application to the $A\sim100$ region
is discussed in section~\ref{s_appli}.
In the final section~\ref{s_conc}
the conclusions and outlook are presented.
Preliminary versions of this work were presented in references~\cite{boyukata08,boyukata10}.

\section{Parametrization of the Hamiltonian}
\label{s_param}
The Hamiltonian used in this paper is of the form
\begin{equation}
\hat H=
\epsilon\,\hat n_d+
\kappa\,\hat Q\cdot\hat Q+
\kappa'\,\hat L\cdot\hat L+
\kappa''\,\hat P_+\cdot\hat P_-+
\lambda\,\hat n_d^2,
\label{e_ham}
\end{equation}
where $\hat n_d$, $\hat P_+$, $\hat L$ and $\hat Q$
are the boson-number, pairing, angular momentum and quadrupole operators,
defined as
\begin{eqnarray}
\hat n_d&=&\sqrt{5}[d^\dag\times\tilde d]^{(0)}_0,
\nonumber\\
\hat P_+&=&[s^\dag\times s^\dag+\sqrt{5}\,d^\dag\times d^\dag]^{(0)}_0,
\qquad
\hat P_-=\left(\hat P_+\right)^\dag,
\nonumber\\
\hat L_\mu&=&\sqrt{10}[d^\dag\times\tilde d]^{(1)}_\mu,
\nonumber\\
\hat Q_\mu&=&[d^\dag\times\tilde s+s^\dag\times\tilde d]^{(2)}_\mu+
\chi[d^\dag\times\tilde d]^{(2)}_\mu.
\end{eqnarray}
Equation~(\ref{e_ham}) defines an \mbox{IBM-1} Hamiltonian
in terms of the six parameters $\epsilon$, $\kappa$, $\kappa'$, $\kappa''$, $\lambda$ and $\chi$.
If one is interested in the properties of a single nucleus,
this is a possible form of the most general Hamiltonian
with the advantage that the first four terms have been used extensively
in phenomenological fits to nuclear spectra~\cite{Iachello87}.
The last term $\hat n_d^2$ gives rise to a so-called `$\tau$-compression'
which increases the moment of inertia with increasing angular momentum
(or with increasing $d$-boson seniority $\tau$)~\cite{Pan92}.
In addition to measured energy spectra,
also electric quadrupole transition rates are considered in the fit.
These are calculated in \mbox{IBM-1}
using the E2 operator $\hat T_\mu({\rm E2})=e_{\rm b}\hat Q_\mu$
where $e_{\rm b}$ is the effective charge of a boson
and with a quadrupole operator
which is consistent with the one used in the Hamiltonian,
following the consistent-$Q$ formalism (CQF)~\cite{Warner82}.

Although reasonable results are obtained with a constant Hamiltonian,
they are considerably improved
if parameters are allowed to vary with the number of nucleons in the valence shell,
as suggested by the shell-model interpretation of the \mbox{IBM-1}.
We propose here a dependence of the generic form
\begin{equation}
x=\sum_{ij}x_{ij}(f_\nu)^i(f_\pi)^j,
\label{e_param}
\end{equation}
where $x$ is a parameter of the Hamiltonian, that is,
$x=\epsilon$, $\kappa$, $\kappa'$, $\kappa''$ or $\lambda$,
and $f_\rho$ is the fractional filling of the neutron ($\rho=\nu$)
or the proton ($\rho=\pi$) valence shell, that is,
$f_\rho\equiv n_\rho/\Omega_\rho$
with $n_\rho$ the number of valence neutrons or protons
and $\Omega_\rho$ the size of the corresponding valence shell.
Note that in the parametrization~(\ref{e_param})
neutrons and protons are always counted
as particles and never as holes.
While microscopic arguments suggest a dependence on
$F_\rho\equiv\bar n_\rho/\Omega_\rho$ rather than on $f_\rho$
(where $\bar n_\rho$ is the number of particles or holes,
whichever is smaller counted from the nearest closed shell),
the latter parametrization has the drawback of yielding spectra
that are symmetric around mid-shell.
Since this particle--hole symmetry
is broken in the nuclei considered here, especially in the Pd isotopes,
we prefer the parametrization~(\ref{e_param}).

\section{Application to the $A\sim100$ region}
\label{s_appli}
In the present application to the $A\sim100$ region
a linear dependence on the fractional fillings
is assumed for the parameters $\kappa$, $\kappa'$ and $\lambda$
while $\epsilon$ and $\kappa''$ are kept constant.
First, energy levels in the known nuclei
$^{100-104}$Zr, $^{94-110}$Mo, $^{98-114}$Ru and $^{102-120}$Pd
are selected and fitted with the \mbox{IBM-1} Hamiltonian~(\ref{e_ham})
by minimizing the root-mean-square (rms) deviation.
In this first step an initial choice is made for $\chi$,
and $\epsilon$, $\kappa$, $\kappa'$, $\kappa''$ and $\lambda$
are fitted to the energy spectra of all nuclei considered in the fit
with no dependence on the number of neutrons or protons.
Next, the linear dependence on the fractional filling of the parameters
$\kappa$, $\kappa'$ and $\lambda$
is determined again by minimizing the rms deviation.
The minimization in this case involves 11 parameters,
three for each coefficient in the Hamiltonian with a fractional-filling dependence
together with $\epsilon$ and $\kappa''$.
Once the wave functions of all states
have been determined in this unified fit,
the E2 transition rates are fitted
by adjusting the value of the boson effective charge $e_{\rm b}$
but keeping to the same value of $\chi$.
Since $B$(E2) properties depend rather strongly on $\chi$,
it is at this point that it can established whether its value is appropriate or not.
The entire procedure is repeated for a new value of $\chi$,
until a reasonable compromise is found
between the results of the energy and E2 fits.

\begin{table}
\caption{Root-mean-square deviations $\Delta$ in the energies and $B$(E2) values
as a function of the quadrupole parameter $\chi$.}
\label{t_chi}
\begin{indented}
\item[]\begin{tabular}{@{}crrrrrrrrr}
\br
$\chi$&
$-0.05$&$-0.1$&$-0.2$&$-0.3$&$-0.4$&$-0.6$&$-0.8$&$-1.0$\\
\mr
$\Delta(E)$~(keV)&126&122&120&122&125&129&131&133\\
$\Delta_1({\rm E2})$~($e^2$b$^2$)&0.26&0.13&0.10&0.11&0.14&0.17&0.15&0.16\\
$\Delta_2({\rm E2})$&3.44&2.29&2.02&2.12&2.34&2.54&2.20&2.21\\
\br
\end{tabular}
\end{indented}
\end{table}
This procedure is illustrated in table~\ref{t_chi}.
The rms deviations in the energies and $B$(E2) values
are quantified with following definitions:
\begin{eqnarray}
\Delta(E)&=&\sqrt{\frac{1}{N_E}\sum_i\left(E_{\rm ex}^i-E_{\rm th}^i\right)^2},
\nonumber\\
\Delta_1({\rm E2})&=&
\sqrt{\frac{1}{N_{\rm E2}}
\sum_i\left(B({\rm E2})_{\rm ex}^i-B({\rm E2})_{\rm th}^i\right)^2},
\nonumber\\
\Delta_2({\rm E2})&=&
\sqrt{\frac{1}{N_{\rm E2}}
\sum_i\left(\frac{B({\rm E2})_{\rm ex}^i-B({\rm E2})_{\rm th}^i}{B({\rm E2})_{\rm ex}^i}\right)^2},
\end{eqnarray}
where the sums are over the available data points,
$N_E$ and $N_{\rm E2}$ in number, respectively.
Note that the first rms deviation for the E2 data, $\Delta_1({\rm E2})$,
is rather insensitive to small E2 transitions
while $\Delta_2({\rm E2})$ probably assigns too much weight  to them.
The rms deviations vary rather weakly
except for small values of $\chi$,
but both energies and E2 transitions
are optimized for $\chi\approx-0.20$.
This value of $\chi$
and the parameters given in table~\ref{t_param}
determine all spectra shown below.
\begin{table}
\caption{Parameters of the Hamiltonian~(\ref{e_ham}) in units of keV.}
\label{t_param}
\begin{indented}
\item[]\begin{tabular}{@{}ccccccccccc}
\br
$\epsilon$&
$\kappa_{00}$&$\kappa_{10}$&$\kappa_{01}$&
$\kappa'_{00}$&$\kappa'_{10}$&$\kappa'_{01}$&
$\kappa''$&
$\lambda_{00}$&$\lambda_{10}$&$\lambda_{01}$\\
\mr
$695$&
$-35.8$&$-72.4$&$47.9$&
$18.3$&$12.8$&$-15.8$&
$1.4$&
$-115.9$&$-166.1$&$161.2$\\
\br
\end{tabular}
\end{indented}
\end{table}

In the above procedure
care has been taken to undertake a gradual release of the parameters,
instead of immediately attempting a full 11-parameter fit.
This method has proven to be numerically stable
but it offers no guarantee that the global minimum
of the 11-dimensional parameter space will be reached,
instead of a local one.
Finding the global minimum is all the more problematic
since some of the parameters,
especially those that account for the fractional-filling dependence,
offer little intuitive insight
and are thus entirely unknown {\it a priori}.
The character of the minimum
corresponding to the parameters of table~\ref{t_param}
has been checked by choosing different starting values
for the $x_{01}$ and $x_{10}$ parameters,
leading always to final parameter sets
that are close to the one shown, within numerical errors.
While this is not a proof of the global character of the minimum,
it is a good indication of it.

\begin{figure}
\begin{center}
\includegraphics[width=16cm]{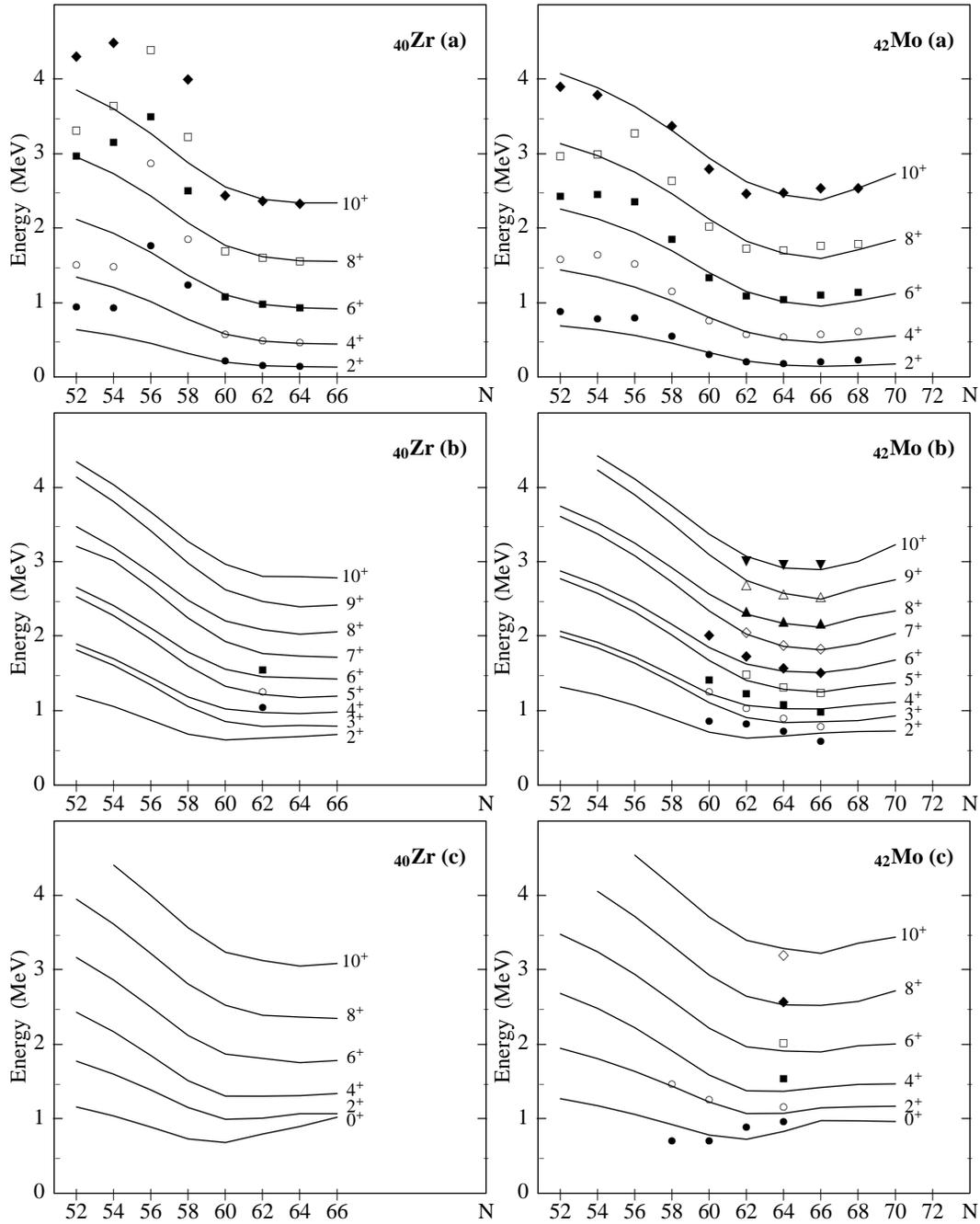}
\end{center}
\caption{
The experimental (symbols) and calculated (lines)
energy spectra for the Zr isotopes with $52\leq N\leq66$ (left)
and the Mo isotopes with $52\leq N\leq70$ (right):
(a) ground-state, (b) quasi-$\gamma$ and (c) quasi-$\beta$ bands.
Only levels in Zr isotopes with $N\geq60$ are fitted.
All experimental levels shown for Mo are included in the fit.}
\label{f_enzrmo}
\end{figure}
\begin{figure}
\begin{center}
\includegraphics[width=16cm]{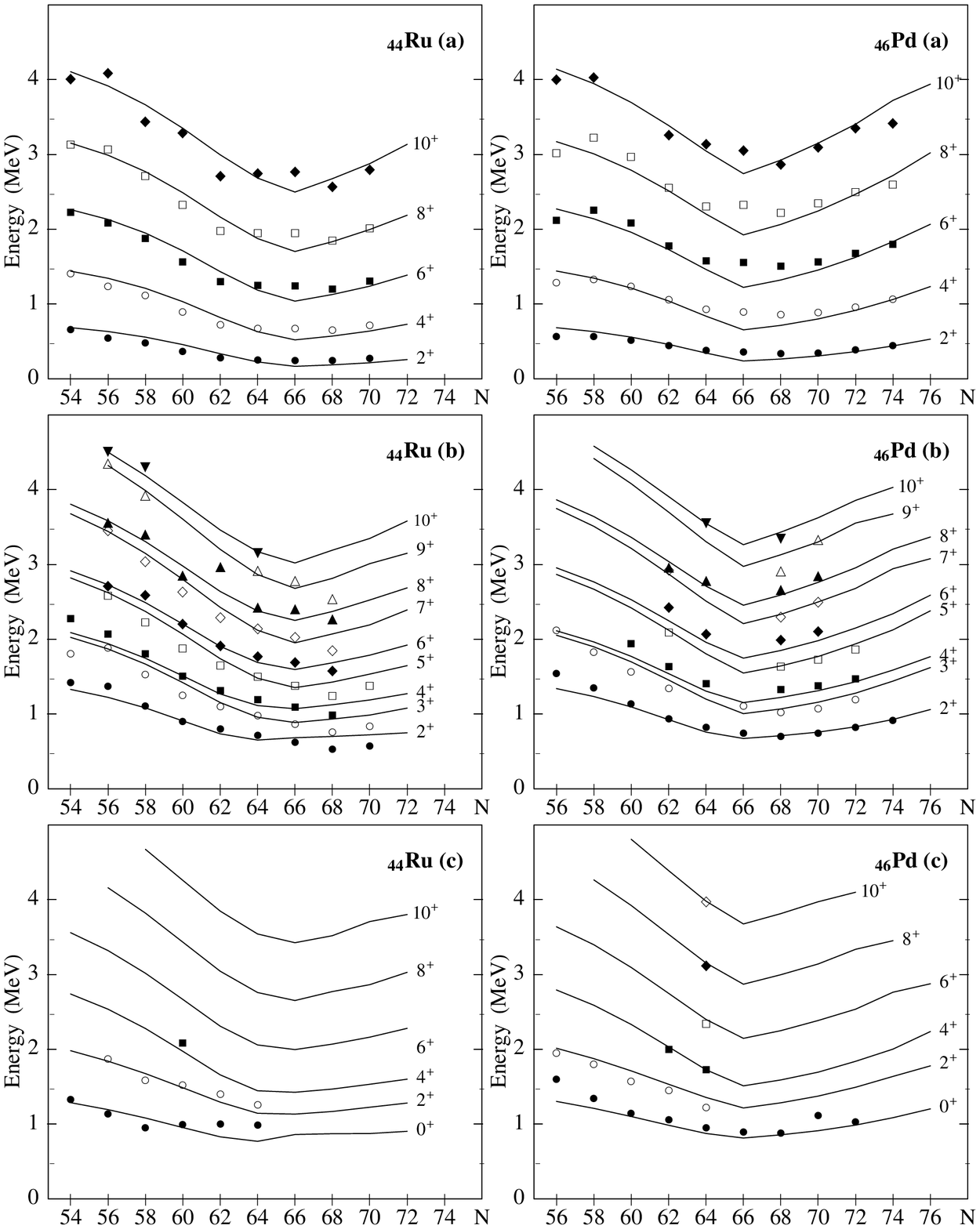}
\end{center}
\caption{
The experimental (symbols) and calculated (lines)
energy spectra for the Ru isotopes with $54\leq N\leq72$ (left)
and for the Pd isotopes with $56\leq N\leq76$ (right):
(a) ground-state, (b) quasi-$\gamma$ and (c) quasi-$\beta$ bands.
All experimental levels shown are included in the fit.}
\label{f_enrupd}
\end{figure}
The experimental and calculated energy spectra
are compared in figures~\ref{f_enzrmo} and \ref{f_enrupd}.
The experimental energies are taken
from the National Nuclear Data Center~\cite{NNDC}
with additional information concerning the nuclei
$^{108}$Pd and $^{112,118,120}$Pd
from the references~\cite{Alcantara05,Stoyer07}.
Some of the spin assignments are uncertain (brackets in NNDC);
these are not shown on the figure to avoid overcrowding.
In the even--even Mo, Ru and Pd isotopes with $N\geq52$,
all known levels of the ground-state, quasi-$\gamma$ and quasi-$\beta$ bands
with angular momentum up to $J=10$ are considered in the fit
while for Zr only isotopes with $N\geq60$ are included.
Figure~\ref{f_enzrmo} nevertheless also shows
the experimental levels for $^{92-98}$Zr
({\em not} included in the fit)
merely to indicate that these strongly deviate
from what is predicted from systematics.
To give an illustration of the deviations,
the $2^+_1$ level in $^{96}$Zr is observed at 1751~keV
while it is calculated at $444$~keV.
These differences are indicative
of the closure of the $1d_{5/2}$ shell at $N=56$,
resulting in a nearly doubly-magic behaviour of $^{96}$Zr.
These considerations can be made quantitative
by including the ground-state bands
of some of the lighter Zr isotopes in the energy fit.
As illustrated in table~\ref{t_zr}
even the inclusion of just one isotope, $^{98}$Zr,
yields a dramatic increase in the rms deviation
of the fit to the entire region.
Effects of $1d_{5/2}$ shell persist in $^{100}$Zr
which exhibits a low-lying $K^\pi=0^+$ band at 331~keV.
This clearly cannot be
the collective first-excited $K^\pi=0^+$ band of the \mbox{IBM-1}
and therefore the band is not included in the fit.
The next-excited $K^\pi=0^+$ band at 829~keV
is possibly the quasi-$\beta$ band of the \mbox{IBM-1},
which is calculated at 682~keV.
\begin{table}
\caption{Root-mean-square deviations $\Delta(E)$
when Zr isotopes with neutron number $N\geq N_0$
are included in the fit.}
\label{t_zr}
\begin{indented}
\item[]\begin{tabular}{@{}crrrrr}
\br
$N_0$&60&58&56&54&52\\
\mr
$\Delta(E)$~(keV)&120&158&214&215&217\\
\br
\end{tabular}
\end{indented}
\end{table}

We emphasize that the \mbox{IBM-1} parametrization~(\ref{e_param})
depends on the counting of the number of {\em valence} nucleons
and hence on the definition of shell closures and magic numbers.
Since $N=56$ is not included as a magic number,
it is therefore no surprise that the structural features
associated with this shell closure
are absent from the present calculations.
The inclusion of {\em all} Zr isotopes in a successful fit
would necessarily require a refined parametrization
that takes account of the $N=56$ closure.
However, even at the present level of approximation,
we believe to have illustrated with this example
that the unified fits of the type used here
are sensitive to nuclear shell effects.
This feature could be of use
to detect precursor effects of shell closures
in extrapolations toward unknown regions of the nuclear chart.

In the present application excluding Zr isotopes with $N<60$,
we arrive at a description of 335 collective levels in 31 nuclei
with a rms deviation from the observed level energies of 120~keV.
The most notable systematic discrepancies are found for
(i) the spectra of $N\sim56$ Mo isotopes
due to some remnant effect of the $N=56$ closure,
(ii) the energy of the $\gamma$ band in $^{102}$Zr
which is predicted lower than it is observed,
and (iii) some mid-shell nuclei like $^{100}$Ru and $^{102}$Pd
which behave less collectively than predicted by \mbox{IBM-1} systematics.

On the basis of the parametrization thus established
it is now possible to make predictions for more exotic nuclei
for which up to now no spectroscopic information is available.
This is illustrated in figures~\ref{f_enzrmo} and \ref{f_enrupd}
where systematically the predictions
for the next (unknown) even--even isotopes are shown,
that is, for $^{106}$Zr, $^{112}$Mo, $^{116}$Ru and $^{122}$Pd.
Clearly, predictions further away from stability are possible as well
but become increasingly uncertain.

\begin{figure}
\begin{center}
\includegraphics[width=14cm]{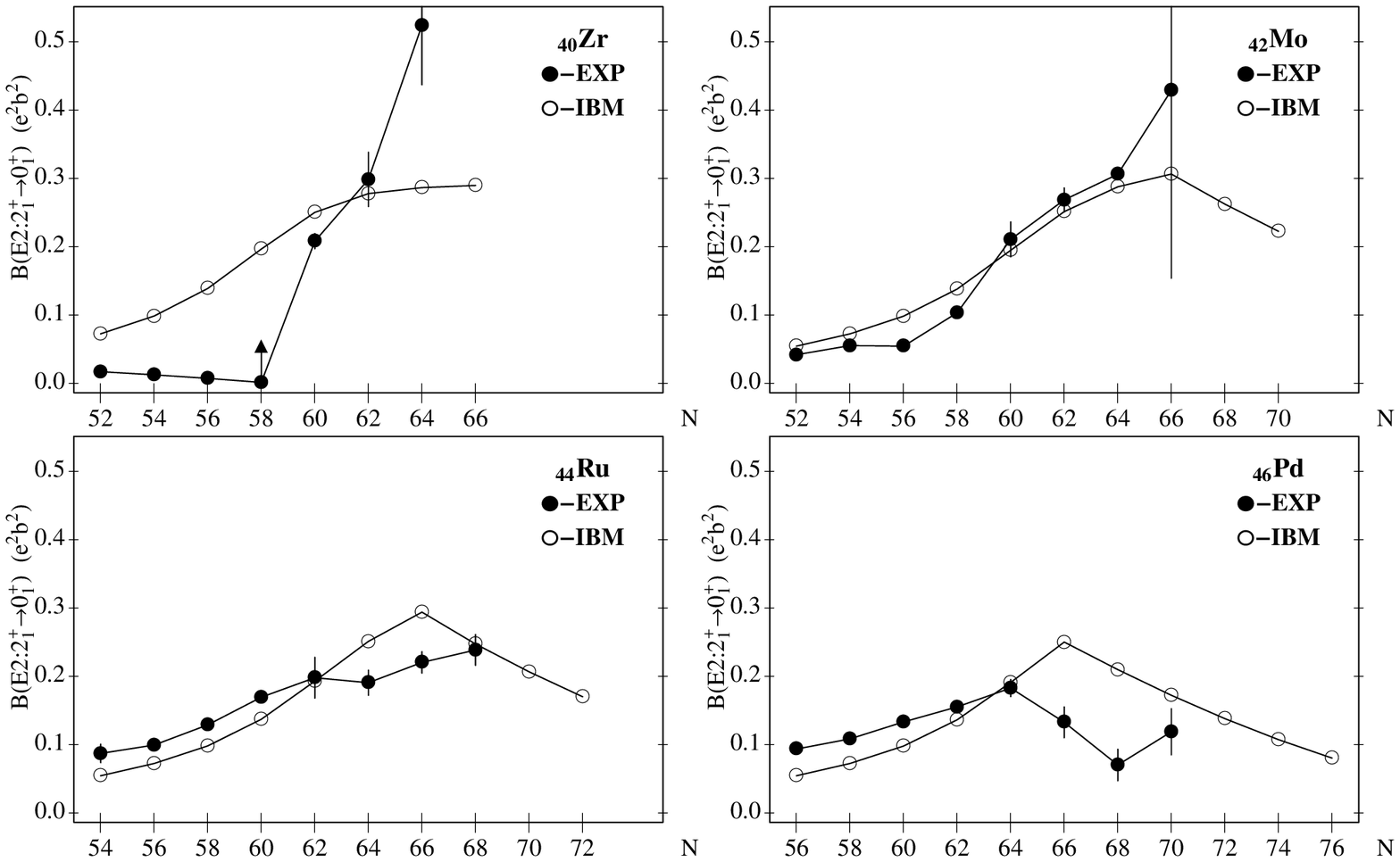}
\end{center}
\caption{The experimental (full symbols)
and calculated (open symbols) $B({\rm E2};2^+_1\rightarrow0^+_1)$ values
in the Zr, Mo, Ru and Pd isotopes.}
\label{f_be20}
\end{figure}
\begin{figure}
\begin{center}
\includegraphics[width=14cm]{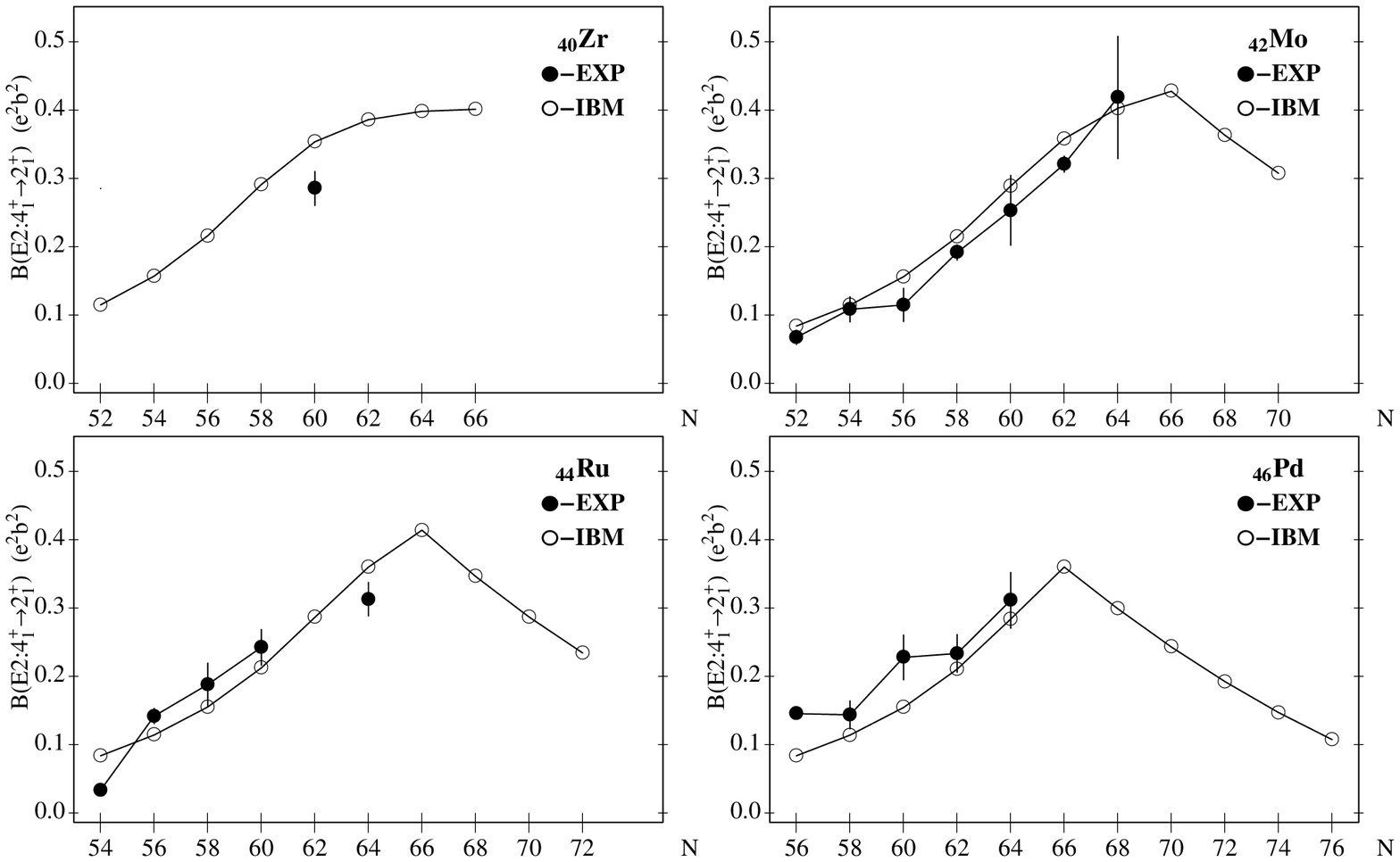}
\end{center}
\caption{The experimental (full symbols)
and calculated (open symbols) $B({\rm E2};4^+_1\rightarrow2^+_1)$ values
in the Zr, Mo, Ru and Pd isotopes.}
\label{f_be42}
\end{figure}
\begin{figure}
\begin{center}
\includegraphics[width=14cm]{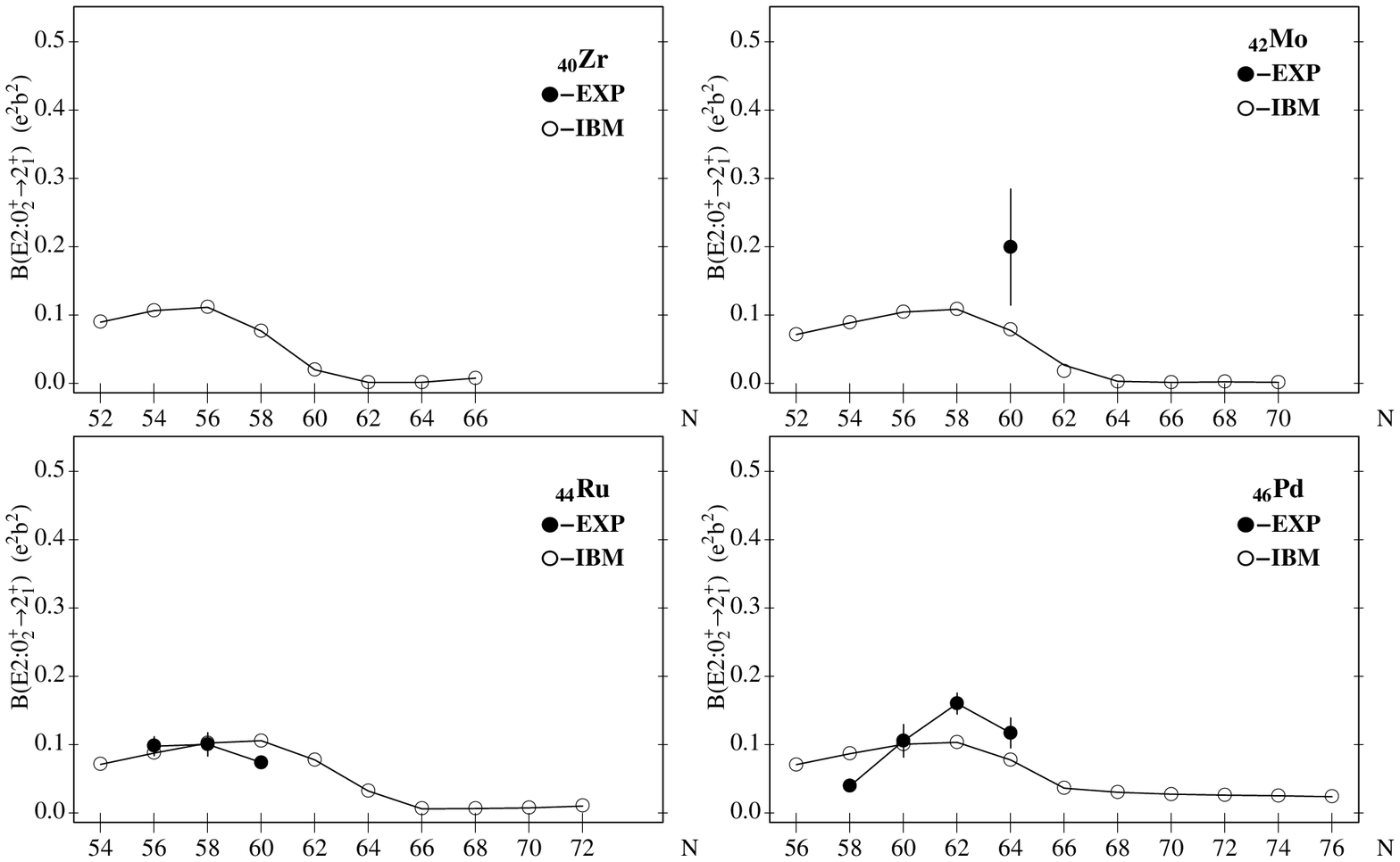}
\end{center}
\caption{The experimental (full symbols)
and calculated (open symbols) $B({\rm E2};0^+_2\rightarrow2^+_1)$ values
in the Zr, Mo, Ru and Pd isotopes.}
\label{f_be02}
\end{figure}
Once the wave functions of all states
have been determined in this unified fit,
one can also calculate the E2 transition rates
with an overall effective charge $e_{\rm b}=0.097$~$e$b,
adjusted to reproduce
the $2^+_1\rightarrow0^+_1$ and $4^+_1\rightarrow2^+_1$ transitions.
The results for the
$B({\rm E2};2^+_1\rightarrow0^+_1)$
and $B({\rm E2};4^+_1\rightarrow2^+_1)$ values
are shown in figures~\ref{f_be20} and \ref{f_be42},
respectively.
These are a measure of the collective behaviour of the nucleus
and our calculation generally follows the observed trends.
A glaring exception to this overall agreement
is the $2^+_1\rightarrow0^+_1$ transition
in the Zr isotopes with $N<60$
where the observed $B$(E2) value
is only fraction from what is expected from systematics.
This again is due to the neglect in the calculation
of the $1d_{5/2}$ closure at $N=56$.
There is also a mid-shell mismatch for the Pd isotopes
which is calculated at $N=66$ but observed at $N=64$.
Figure~\ref{f_be02} shows the $B({\rm E2};0^+_2\rightarrow2^+_1)$ values.
In the vibrational U(5) limit this transition corresponds to one
from a two-phonon to a one-phonon state which is allowed
while in the more deformed limits SO(6) and SU(3)
this $B$(E2) value is much smaller.
This argument gives a qualitative understanding
of the theoretical results shown in figure~\ref{f_be02}.

\begin{table}
\caption{ Experimental and calculated
$B({\rm E2};J^\pi_{\rm i}\rightarrow J^\pi_{\rm f})$ values
in units of $10^2$ e$^2$fm$^4$
in $^{100}$Ru, $^{104}$Pd and $^{108}$Pd.}
\label{t_e2}
\begin{indented}
\item[]\begin{tabular}{@{}ccrrcrrcrr}
\br
&~~~&\multicolumn{2}{c}{$^{100}$Ru}
&~~~&\multicolumn{2}{c}{$^{104}$Pd}
&~~~&\multicolumn{2}{c}{$^{108}$Pd}\\
\cline{3-4}\cline{6-7}\cline{9-10}
$J^\pi_{\rm i}\rightarrow J^\pi_{\rm f}$
&&Expt&IBM-1&&Expt&IBM-1&&Expt&IBM-1\\
\mr
$2_1^+\rightarrow 0_1^+$&&9.8~(0.1)     &7.1    &&11~(0.6)  &7.1  &&15~(0.5)  &14\\
$4_1^+\rightarrow 2_1^+$&&14~(1.1)      &11     &&14~(2)    &11   &&23~(2.8)  &21\\
$6_1^+\rightarrow 4_1^+$&&$<47$         &13     &&---       &13   &&33~(3.4)  &24\\
$8_1^+\rightarrow 6_1^+$&&---           &12     &&---       &12   &&45~(6.1)  &25\\
$0_2^+\rightarrow 2_1^+$&&9.7~(1.4)     &8.6    &&3.8~(0.4) &8.5  &&16~(1.5)  &10\\
$2_2^+\rightarrow 0_1^+$&&$0.52(^{\rm +0.11}_{\rm -0.14})$&0.004  &&0.38~(0.03)&0.004&&0.25~(0.02)&0.10\\
$2_2^+\rightarrow 2_1^+$&&8.5~(0.11)    &11     &&6.3~(4.9) &11   &&22~(1.5)  &19\\
$4_2^+\rightarrow 2_1^+$&&$0.52(^{\rm +0.22}_{\rm -0.06})$&0.003&&0.17~(0.17) &0.003&&0.04~(0.04)  &0.04\\
$4_2^+\rightarrow 2_2^+$&&---           &6.7    &&7.3~(7.3) &6.7  &&17~(1.8)  &12\\
$4_2^+\rightarrow 4_1^+$&&7.4~(4.7)     &6.0    &&2.9~(2.9) &6.0  &&9.2~(1.8) &10\\
\br
\end{tabular}
\end{indented}
\end{table}
As a further illustration of the kind of accuracy
obtained for the $B$(E2) values in this overall fit,
we show in table~\ref{t_e2} the results for three nuclei,
$^{100}$Ru, $^{104}$Pd and $^{108}$Pd,
for which extensive E2 data is known.

\begin{figure}
\begin{center}
\includegraphics[width=14cm]{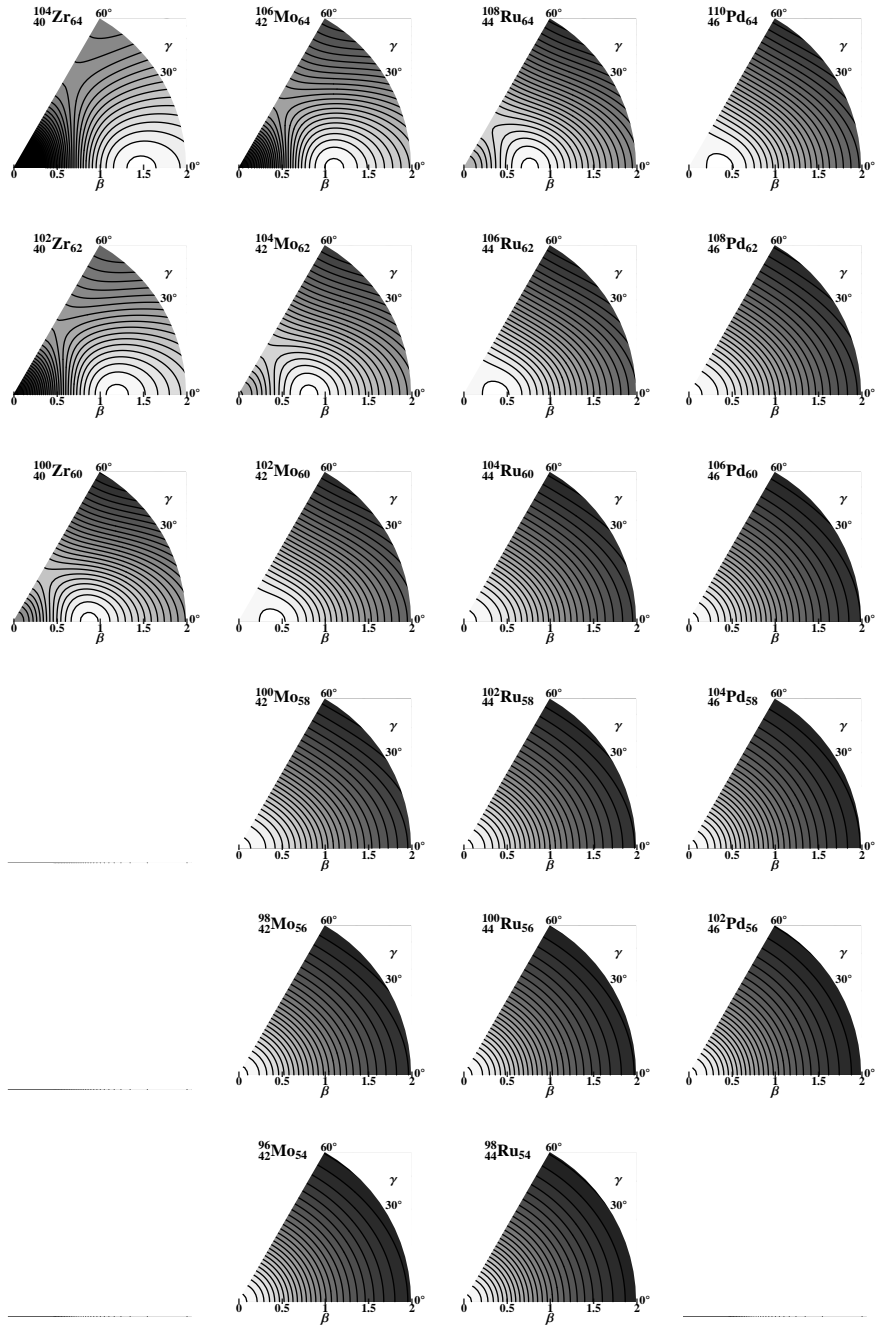}
\end{center}
\caption{Potential energy surfaces for all nuclei (first part).}
\label{f_pes1}
\end{figure}
\begin{figure}
\begin{center}
\includegraphics[width=14cm]{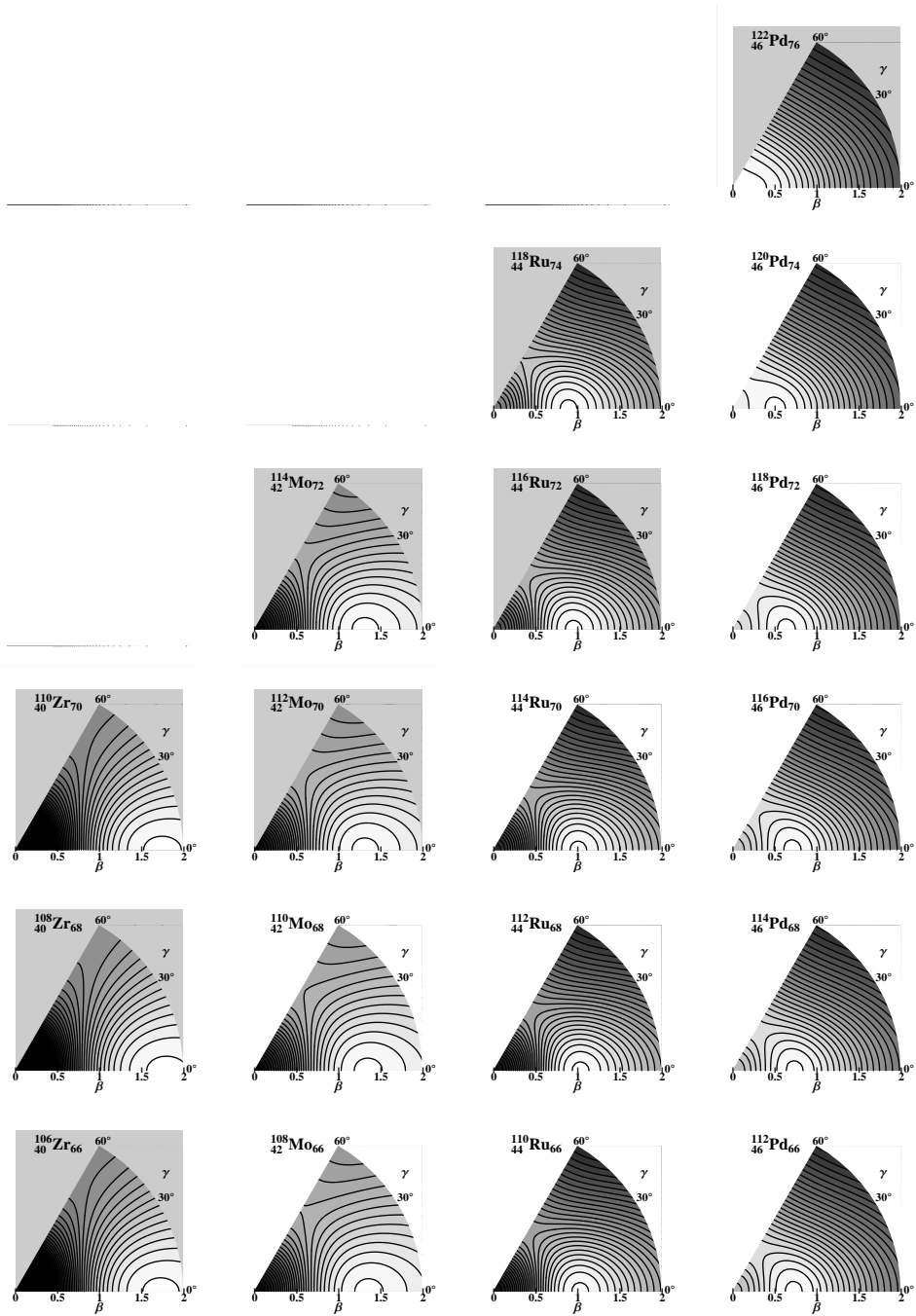}
\end{center}
\caption{Potential energy surfaces for all nuclei (second part). The
grey background indicates a potential surface for a predicted
nucleus.} \label{f_pes2}
\end{figure}
The geometric character of all nuclei
can be visualized by plotting the potential energy surface $V(\beta,\gamma)$
obtained from the \mbox{IBM-1} Hamiltonian
in the classical limit~\cite{Ginocchio80,Dieperink80,Bohr80}.
The classical limit of the \mbox{IBM-1} Hamiltonian~(\ref{e_ham}) is
\begin{eqnarray}
V(\beta,\gamma)&=&
\frac{N}{1+\beta^2}
\left(a^{(1)}_{00}+
a^{(1)}_{10}\beta^2\right)+
\nonumber\\&&
\frac{N(N-1)}{(1+\beta^2)^2}
\left(a^{(2)}_{00}+
a^{(2)}_{10}\beta^2+
a^{(2)}_{01}\beta^3\cos3\gamma+
a^{(2)}_{20}\beta^4\right),
\label{e_climit}
\end{eqnarray}
where the coefficients $a^{(n)}_{kl}$ are given by
\begin{eqnarray}&&
a^{(1)}_{00}=5\kappa,
\quad
a^{(1)}_{10}=\epsilon+(1+\chi^2)\kappa+6\kappa'+\lambda,
\quad
a^{(2)}_{00}={\frac 1 4}\kappa'',
\nonumber\\&&
a^{(2)}_{10}=4\kappa-{\frac 1 2}\kappa'',
\quad
a^{(2)}_{01}=-4\sqrt{\frac 2 7}\chi\kappa,
\quad
a^{(2)}_{20}={\frac 2 7}\chi^2\kappa+{\frac 1 4}\kappa''+\lambda.
\label{e_coeff}
\end{eqnarray}
Since the parameters in this potential
depend on the fractional fillings of the valence shells,
we arrive at a geometry for each nucleus
which changes with neutron and proton numbers
and which has been obtained in an unbiased and straightforward way
from the observed excitation spectra.
This is illustrated in figures~\ref{f_pes1} and~\ref{f_pes2},
where the potentials are shown for all nuclei included in the fit
as well as for some which are predicted.
For the sake of a correct interpretation of these potential surfaces,
we recall that $\gamma$ is identical to
the corresponding parameter in the Bohr--Mottelson geometric model~\cite{BM75}
while $\beta$ is only proportional to it,
the proportionality factor depending on the ratio
of the valence over the total nucleon numbers~\cite{Ginocchio80b}.
The potential energy surface of the nucleus $^{108}$Ru
has recently been calculated in variety of ways.
M\"oller {\it et al.}~\cite{Moeller06} showed
that deviations from axial symmetry can be important
and, in the specific case of $^{108}$Ru,
obtained a potential energy surface with triaxial minimum.
With the simple parametrization of the \mbox{IBM-1} Hamiltonian
as proposed here which is limited to two-body interactions,
no triaxial minimum can be obtained~\cite{Isacker81},
a result confirmed by the surfaces shown
in figures~\ref{f_pes1} and~\ref{f_pes2}.
It is also seen, however, that the specific potential for $^{108}$Ru
is rather soft in $\gamma$.
Triaxial minima can be obtained when cubic terms
are added to the \mbox{IBM-1} Hamiltonian~\cite{Stefanescu07,Sorgunlu08}.
In comparison with the results
obtained in reference~\cite{Sorgunlu08} for the nuclei $^{108,110,112}$Ru,
we observe that the potential surfaces shown here
show less variation with neutron number
since they are derived from a global fit to many nuclei
while in the former study the fits were carried out nucleus by nucleus.
Also shown in figures~\ref{f_pes1} and~\ref{f_pes2}
are the potential energy surfaces for the predicted neutron-rich nuclei.
As can be seen from the figures,
their geometry varies from strongly deformed in the neutron-rich Zr isotopes
to spherical in $^{122}$Pd.

\section{Conclusions}
\label{s_conc}
We have suggested in this paper the use of \mbox{IBM-1}
for global calculations of large regions of the nuclear chart
and have illustrated the idea with an application to the $A\sim100$ region.
The present work calls for several features to be studied in more detail.
The first concerns the technical issue of finding the set of parameters
that corresponds to the {\em global} minimum of the rms deviation.
The minimization algorithm that has been used here
is a straightforward linearization procedure of the eigenvalue problem
associated with the Hamiltonian~(\ref{e_ham})
which has no trouble in finding a {\em local} minimum.
We have found no dependence of the converged parameters
on various choices of their initial values
but it would be of interest to confirm the global character of the minimum
using a different algorithm which samples large parameter regions.
Secondly, the dependence of the parameters
on the fractional fillings $f_\rho$ of the valence neutron and proton shells
requires further study.
We have proposed here a dependence on $n_\rho/\Omega_\rho$
but other choices should be investigated as well,
such as a dependence on $N_\rho/\Omega_\rho$
(where $N_\rho$ is the boson number)
or on Casten's factor $P=N_\nu N_\pi/(N_\nu+N_\pi)$~\cite{Casten85}.
Furthermore, a possible dependence on subshells must also be studied.
A proper treatment of these effects
and their dependence on valence nucleon number
must draw inspiration from the microscopic foundation of the IBM.
For example, in this study we have taken a constant value of $\chi$,
the parameter appearing in the quadrupole operator.
This is possibly a reasonable approximation
for the restricted set of nuclei considered here
but it will certainly be inadequate
if large regions of the nuclear chart are fitted simultaneously.
This inadequacy might be satisfactorily resolved
by a proper estimate of this parameter
based on the shell-model interpretation of the IBM.
An alternative strategy is to adopt a functional dependence of $\chi$
on valence-particle number suggested by microscopy,
containing a few parameters that are fitted to the data.
Finally, we point out that the type of calculation described here
opens up the possibility for a simultaneous and global treatment
of nuclear ground-state properties (such as masses and radii)
in addition to the excited-state properties discussed in this paper.
A possible strategy for merging the calculations
of ground- and excited-state energies
in the framework of the interacting boson model
was outlined in reference~\cite{Isacker09}
with preliminary results for even--even nuclei
in the major shell with $82<N<126$ and $50<Z<82$
but further investigations of this approach are required.

\section*{Acknowledgements}
This work was supported
by the Scientific and Technological Research Council of Turkey (TUBITAK)
under project nr 107T557,
and by the Agence Nationale de Recherche, France,
under contract nr ANR-07-BLAN-0256-03.


\section{References}
\begin{thebibliography}{99}

\bibitem{Bender03}
Bender~M, Heenen~P-H and Reinhard~P-G 2003
{\it Rev.\ Mod.\ Phys.\ } {\bf75} 121

\bibitem{Caurier05}
Caurier~E, Mart\'\i nez-Pinedo~G, Nowacki~F, Poves~A and Zuker~A~P 2005
{\it Rev.\ Mod.\ Phys.\ } {\bf77} 427

\bibitem{Arima76}
Arima~A and Iachello~F 1976
{\it Ann.\ Phys.\ (N.Y.)} {\bf99} 253

\bibitem{Arima78}
Arima~A and Iachello~F 1978
{\it Ann.\ Phys.\ (N.Y.)} {\bf111} 201

\bibitem{Arima79}
Arima~A and Iachello~F 1979
{\it Ann.\ Phys.\ (N.Y.)} {\bf123} 468

\bibitem{Arima77}
Arima~A, Otsuka~T, Iachello~F and Talmi~I 1977
{\it Phys.\ Lett.\ B} {\bf66} 205

\bibitem{Otsuka78}
Otsuka~T, Arima~A and Iachello~F 1978
{\it Nucl.\ Phys.\ A} {\bf309} 1

\bibitem{Iachello87}
Iachello~F and Arima~A 1987
{\it The Interacting Boson Model}
(Cambridge: Cambridge University Press)

\bibitem{Gomez95}
G\'omez~A, Casta\~nos~O and Frank~A 1995
{\it Nucl.\ Phys.\ A} {\bf589} 267

\bibitem{Lalkovski09}
Lalkovski~S and Van~Isacker~P 2009
{\it Phys.\ Rev.\ C} {\bf79} 044307

\bibitem{Casten85b}
Casten~R~F, Frank~W and von~Brentano~P 1985
{\it Nucl.\ Phys.\ A} {\bf444} 133

\bibitem{Casten85}
Casten~R~F 1985
{\it Nucl.\ Phys.\ A} {\bf443} 1

\bibitem{Nomura08}
Nomura~K, Shimizu~N and Otsuka~T 2008
{\it Phys.\ Rev.\ Lett.\ } {\bf101} 142501

\bibitem{boyukata08}
B\"oy\"ukata~M, Van~Isacker~P and Uluer~\.{I} 2008 {\it AIP Conf.\
Proc.\ } {\bf1072} 223

\bibitem{boyukata10}
B\"oy\"ukata~M and Uluer~\.{I} 2010 {\it AIP Conf.\ Proc.\ }
{\bf1231} 201

\bibitem{Pan92}
Pan~X-W, Otsuka~T, Chen~J-Q and Arima~A 1992
{\it Phys.\ Lett.\ B} {\bf287} 1

\bibitem{Warner82}
Warner~D~D and Casten~R~F 1982
{\it Phys.\ Rev.\ Lett.\ } {\bf48} 1385

\bibitem{NNDC}
National Nuclear Data Center (NNDC)
http://www.nndc.bnl.gov/


\bibitem{Alcantara05}
Alc\'antara-N\'u\~nez~J~A, Oliveira~J~R~B, Cybulska~E~W, Medina~N~H,
Rao~M~N, Ribas~R~V, Rizzutto~M~A, Seale~W~A, Falla-Sotela~F and Wiedemann~K~T 2005
{\it Phys.\ Rev.\ C} {\bf71} 054315

\bibitem{Stoyer07}
Stoyer~M~A {\it et al} 2007
{\it Nucl.\ Phys.\ A} {\bf787} 455c

\bibitem{Ginocchio80}
Ginocchio~J~N and Kirson~M~W 1980
{\it Phys.\ Rev.\ Lett.\ } {\bf44} 1744

\bibitem{Dieperink80}
Dieperink~A~E~L, Scholten~O and Iachello~F 1980
{\it Phys.\ Rev.\ Lett.\ } {\bf44} 1747

\bibitem{Bohr80}
Bohr~A and Mottelson~B~R 1980
{\it Phys.\ Scripta} {\bf22} 468

\bibitem{BM75}
Bohr~A and Mottelson~B~R 1975
{\it Nuclear Structure. II Nuclear Deformations} (Benjamin: New York)

\bibitem{Ginocchio80b}
Ginocchio~J~N and Kirson~M~W 1980
{\it Nucl.\ Phys.\ A} {\bf350} 31

\bibitem{Moeller06}
M\"oller~P, Bengtsson~R, Carlsson~B~G, Olivius~P and Ichikawa~T 2006
{\it Phys.\ Rev.\ Lett.\ } {\bf97} 162502

\bibitem{Isacker81}
Van~Isacker~P and Chen~J-Q 1981
{\it Phys.\ Rev.\ C} {\bf24} 684

\bibitem{Stefanescu07}
Stefanescu~I, Gelberg~A, Jolie~J, Van~Isacker~P, von~Brentano~P,
Luo~Y~X, Zhu~S~J, Rasmussen~J~O, Hamilton~J~H, Ramayya~A~V and Che~X~L 2007
{\it Nucl.\ Phys.\  A} {\bf789} 125

\bibitem{Sorgunlu08}
Sorgunlu~B and Van~Isacker~P 2008
{\it Nucl.\ Phys.\ A} {\bf808} 27

\bibitem{Isacker09}
Van~Isacker~P 2009
{\it Rev.\ Mex.\ F\'\i s.\ }{\bf55} (2) 66

\end{thebibliography}
\end{document}